\documentclass[twocolumn,prd,footinbib,aps,prl,floats,floatfix,amsmath,superscriptaddress,
amssymb,longbibliography,secnumarabic]{revtex4-1} %
\usepackage[final]{graphicx}
\usepackage{hyperref}
\usepackage{amsmath}
\usepackage{bbm}
\usepackage{amsfonts}
\usepackage{amssymb}
\usepackage{latexsym}
\usepackage{graphicx}
\usepackage[english]{babel}
\usepackage{multirow}
\usepackage{float}
\usepackage{url}
\usepackage{slashed}
\usepackage{xcolor} 
\usepackage[utf8]{inputenc}
\usepackage{verbatim}
\usepackage{stackengine}

\def\sfrac#1#2{{\textstyle{#1\over #2}}}
\newcommand{\be}{\begin{equation}}
\newcommand{\ee}{\end{equation}}
\newcommand{\ba}{\begin{array}}
\newcommand{\ea}{\end{array}}
\newcommand{\bea}{\begin{eqnarray}}
\newcommand{\eea}{\end{eqnarray}}
\newcommand{\sss}{\scriptscriptstyle}

\newcommand{\nn}{\nonumber}

\newcommand{\diff}{\mathrm{{d}}}

\newcommand{\new}[1]{{\color{red} #1}}

\usepackage[utf8]{inputenc}

\begin{document}
\rightline{CERN-TH-2024-016}
\title{Self-interacting dark matter solves the final parsec problem \\ of supermassive black hole mergers}

\author{Gonzalo Alonso-{\'A}lvarez}
\email{gonzalo.alonso@utoronto.ca}
\thanks{ORCID: \href{https://orcid.org/0000-0002-5206-1177}{0000-0002-5206-1177}}
\affiliation{Department of Physics, University of Toronto, Toronto, ON M5S 1A7, Canada}
\affiliation{McGill University Department of Physics \& Trottier Space Institute, 3600 Rue University, Montr\'eal, QC, H3A 2T8, Canada}
\author{James M.\ Cline}
\email{jcline@physics.mcgill.ca}
\thanks{ORCID: \href{https://orcid.org/0000-0001-7437-4193}{0000-0001-7437-4193}}
\affiliation{McGill University Department of Physics \& Trottier Space Institute, 3600 Rue University, Montr\'eal, QC, H3A 2T8, Canada}
\affiliation{CERN, Theoretical Physics Department, Geneva, Switzerland}
\author{Caitlyn Dewar}
\email{caitlyn.dewar@mail.mcgill.ca}
\thanks{ORCID: \href{https://orcid.org/0009-0002-2191-7224}{0009-0002-2191-7224}}
\affiliation{McGill University Department of Physics \& Trottier Space Institute, 3600 Rue University, Montr\'eal, QC, H3A 2T8, Canada}

\begin{abstract}
Evidence for a stochastic gravitational wave (GW) background, plausibly originating from the merger of supermassive black holes (SMBHs), is accumulating with observations from pulsar timing arrays. 
An outstanding question is how inspiraling SMBHs get past the ``final parsec'' of separation, where they have a tendency to stall before GW emission alone can make the binary coalesce.  We argue that dynamical friction from the dark matter (DM) spike surrounding the black holes is sufficient to resolve this puzzle, if the
DM has a 
self-interaction cross section of order
cm$^2$/g. The same effect leads to a softening of the
GW spectrum at low frequencies as suggested by the current data.  For collisionless cold DM, the friction deposits so much energy that the spike is disrupted and cannot bridge the final parsec, while for self-interacting DM, the isothermal core of the halo can act as a reservoir for the energy liberated from the SMBH orbits.  A realistic velocity dependence, such as generated by the exchange of a massive mediator like a dark photon, is favored to give a good fit to the GW spectrum while providing a large enough core. A similar velocity dependence has been advocated for solving the small-scale structure problems of cold DM.

\end{abstract} 
\maketitle

{\bf 1. Introduction.}
Nearly 40 years after its theoretical basis was established
\cite{Hellings:1983fr}, gravitational wave astronomy has entered a new era with the advent of pulsar timing arrays.
Differential time delays of pulsar signals, consistent with a stochastic gravitational wave (GW) background at nanoHertz frequencies, were detected in 2021 by NANOGrav~\cite{Aggarwal:2018mgp}, the Parkes Pulsar Timing Array (PPTA)~\cite{Goncharov:2021oub} and 
the European Pulsar Timing Array (EPTA)~\cite{EPTA:2021crs}.
The GW interpretation has been reinforced by the $\sim 3\sigma$ evidence for Hellings-Downs correlations in the
NANOGrav's recent 15-year data analysis~\cite{NANOGrav:2023gor}, and the compatible measurements by PPTA~\cite{Reardon:2023gzh}, EPTA~\cite{EPTA:2023fyk}, and the Chinese Pulsar Timing Array~\cite{Xu:2023wog}.
A plausible origin for the signal are the mergers of supermassive black holes (SMBHs)~\cite{Begelman:1980vb,NANOGrav:2023hfp,EPTA:2023xxk,Ellis:2023dgf} across cosmic time.

One challenge to the SMBH interpretation of the nHz GW background is that the simplest models (assuming GW emission is the only source of energy loss) predict that the timescale for merging once the SMBH separation is of order
1\,pc is larger than a Hubble time; this ``final parsec'' problem suggests that the SMBHs would never merge~\cite{Milosavljevic:2002bn}.
At larger distances, three-body interactions with stars allow the SMBH pair to lose
energy, ``hardening'' the binary and driving the inspiral.
It was suggested that axisymmetry of the galactic halo profile is sufficient to overcome this problem \cite{Khan:2013wbx}, but this has been debated \cite{Vasiliev:2013nha}.  Another possibility is that interactions of the SMBHs with an accretion disk accelerate the infall~\cite{Kocsis:2010xa,Goicovic:2016dul,Goicovic:2018xxi}. 
The simulations in~\cite{Kelley:2016gse} show that these astrophysical mechanisms are generally ineffective to reduce the inspiral time below several Gyr, adding motivation to look for others.

A less-explored mechanism for accelerating the infall is the dynamical friction (DF) \cite{Chandrasekhar:1943ys} experienced by the SMBH pair as it rotates through the surrounding dark matter (DM) halo.  This effect has been studied for ultralight DM~\cite{Vicente:2022ivh,Bromley:2023yfi,Berezhiani:2023vlo,Boudon:2023qbu,Chan:2022gqd,Mitra:2023sny,Aurrekoetxea:2023jwk} and in the context of intermediate- or stellar-mass BH binaries \cite{Eda:2014kra,Tang:2020jhx,Kavanagh:2020cfn,Coogan:2021uqv,Kadota:2023wlm}.
Black holes accumulate surrounding DM overdensities, known as ``spikes''~\cite{Gondolo:1999ef}, which can exceed the galactic DM halo density and enhance the DF damping the BH orbital motion.

\begin{figure}[t]
\centerline{\includegraphics[width=0.95 \columnwidth]{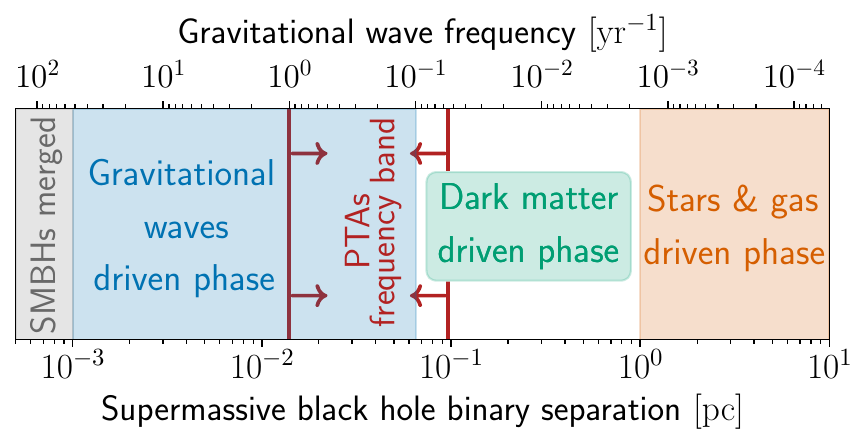}}
\vspace{-0.2cm}
\caption{Agents driving the SMBH binary hardening versus separation (bottom axis) or gravitational wave emission frequency (top axis), compared to the bandwidth of current pulsar timing arrays.
Regions correspond to a typical merger of two $3\times 10^9\,M_\odot$ SMBHs at $z=0$ within an SIDM spike.
}
\label{fig:sketch}
\end{figure}

Some effects of collisionless cold dark matter (CDM) friction were recently considered for SMBH contributions to PTA signals in Ref.~\cite{Shen:2023pan}.
It was shown that the low-frequency turnover in the spectrum, suggested by the data, can be ascribed to DM frictional energy loss, which dominates over GW losses at intermediate BH separations. 
The effect of eccentricity of the SMBH orbits was studied in 
Ref.~\cite{Hu:2023oiu}.
The impact on the final parsec problem has however not yet been addressed. 
Here we show that DM friction drives the binary infall at intermediate separations (see Fig.~\ref{fig:sketch}) and can reduce the inspiral timescale to $\lesssim 1$~Gyr, provided that the DM spike is able to absorb the frictional energy without being disrupted. We argue that this is possible for self-interacting dark matter (SIDM), but not for standard CDM.   

Self-interacting dark matter (SIDM) has been proposed to address discrepancies between the predictions of CDM and observations of galactic structure on small scales, notably the core versus cusp problem (see~\cite{Tulin:2017ara} for a review).
We consider a range of velocity-dependent scattering cross sections, motivated by evidence for scale-dependence of halo cores \cite{Kaplinghat:2015aga},
finding that simple power laws do not optimally fit the GW signal. 
However, a broken power law, as results from a realistic massive force carrier like a dark photon, is able to reproduce the observations, including the 
hint of reduced power at low frequencies.
Our preferred cross section values are compatible with those favored by small-scale structure.

\medskip
{\bf 2. DM density profiles.} The GW signal from merging SMBHs depends upon their masses $M_1$ and $M_2$, parameterized by $q \equiv M_2/M_1 \le 1$, with final SMBH mass $M_\bullet = M_1(1+q)$.
For a given $M_\bullet$ at redshift $z$, we determine the mean values of the Navarro-Frenk-White (NFW)~\cite{Navarro:1995iw} halo parameters of the DM density profile following~\cite{Girelli:2020goz,Kormendy:2013dxa,Klypin:2014kpa}, as described in Appendix \ref{app:halo}.
This provides a starting point for determining the DM spike around the SMBH.

For CDM, the NFW profile is superseded by the DM spike contribution at radii $r < r_{\rm sp}$, with 
spike radius $r_{\rm sp}\cong 0.2 \, r_{\mathrm{2M}}$, where $r_{\mathrm{2M}}$ is the radius at which the mass enclosed by the NFW profile is $2M_\bullet$~\cite{Merritt:2003qc,Merritt:2003qk}. 
For an NFW halo, $r_{\mathrm{2M}}^2 \cong M_\bullet/\pi\rho_s r_s$, assuming $r_{\rm 2M}\ll r_s$.
The spike profile has the form
\begin{equation} \label{eq:spike_profile}
    \rho_{\rm sp}(r) = \rho_{\rm sp}\left({r_{\rm sp}/ r}\right)^{\gamma},
\end{equation}
where $\rho_{\rm sp} = \rho_{\sss \rm NFW}(r_{\rm sp})$, and the exponent $\gamma$ is subject to astrophysical uncertainties~\cite{Ullio:2001fb,Merritt:2002vj,Gnedin:2003rj}, including evolution during the merger \cite{Milosavljevic:2001vi}.
We thus consider $\gamma$ as a free parameter, with physically motivated values varying between $7/3$ for an adiabatically grown spike~\cite{Gondolo:1999ef} and $1/2$ for a spike formed right after a galaxy merger~\cite{Ullio:2001fb}.
The spike can be tapered off by DM annihilations~\cite{Vasiliev:2007vh,Shapiro:2016ypb},
but this will not play a role here.

\begin{figure}[t]
\centerline{\!\!\!\!\!\!\!\!\!\!
\includegraphics[width=0.95 \columnwidth]{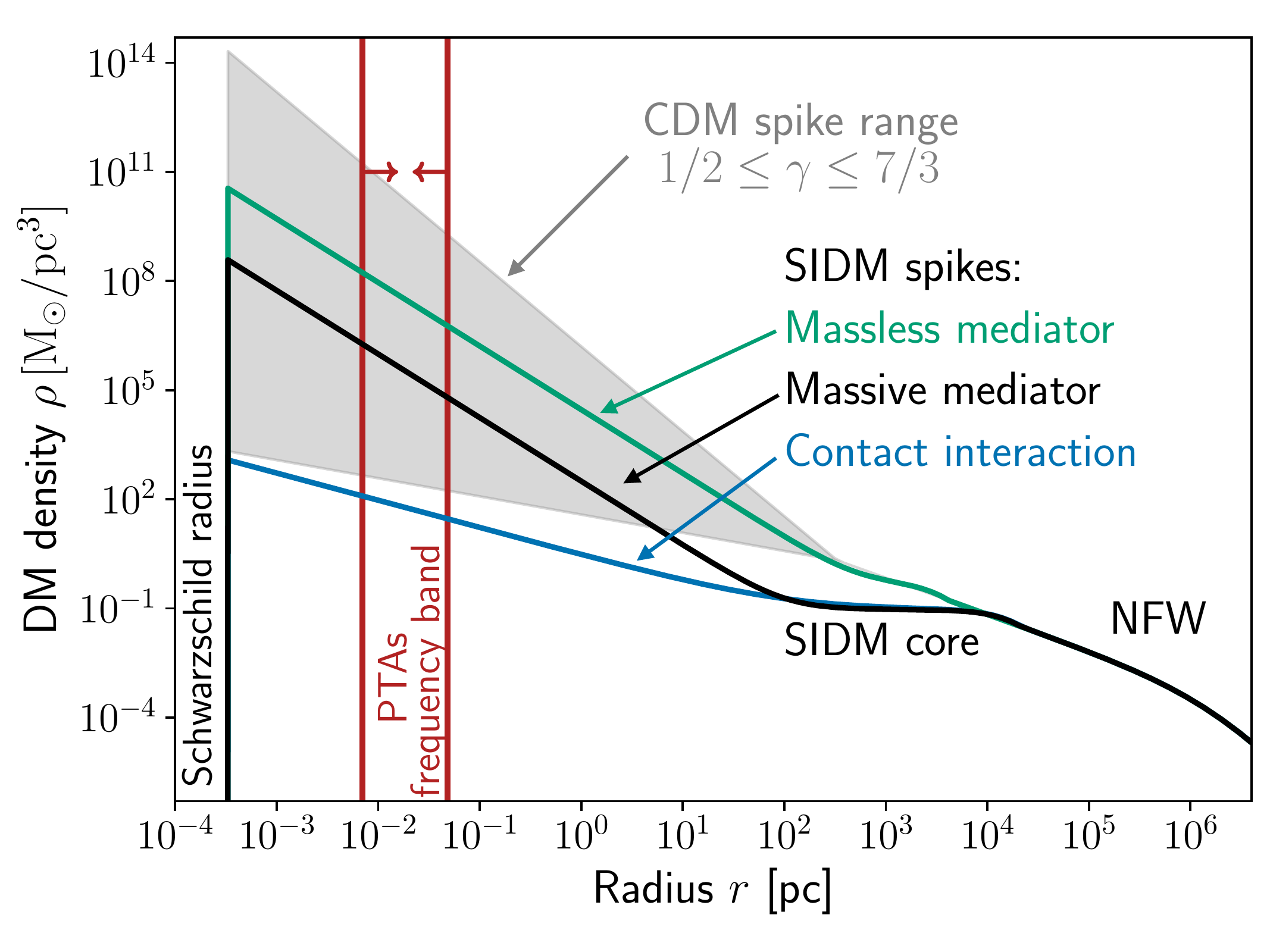}}
\vspace{-0.3cm}
\caption{Spike profiles around a SMBH with $M_\bullet=6\times10^{9}\,M_\odot$ at $z=0$.
\new{NFW parameters of the host galaxy are $r_s\simeq 2~\mathrm{Mpc}$, $\rho_s\simeq3\times 10^{14}\,M_\odot/\mathrm{Mpc}^3$, and $M_{200}\simeq 2\times 10^{16}\,M_\odot$.}
A range of possible CDM spikes is shaded in grey.
For SIDM, the blue (green) line corresponds to a contact interaction (massless mediator) with $\sigma_0/m=3\,\mathrm{cm}^2/\mathrm{g}\  (30\,\mathrm{cm}^2/\mathrm{g}$).
The black line represents a massive mediator with
$\sigma_0/m=3\,\mathrm{cm}^2/\mathrm{g}$ and a \new{transition} velocity  $v_t=500\,\mathrm{km}/\mathrm{s}$.
The age of the core is \new{100\,Myr}.
Vertical red lines delimit the range for which GW emission is detectable at current PTAs.
\new{The profiles are cut off at twice the Schwarzschild radius of the BH \cite{Shapiro:2014oha}.}}
\label{fig:DM_spike}
\end{figure}

For SIDM, the galactic halo is flattened within a distance $r_1$, the radius of the core, due to SI-driven thermalization~\cite{Kaplinghat:2015aga}.
The interaction cross section $\sigma$ determines
$r_1$ by demanding at least one scattering per DM particle for $r<r_1$ during the age of the core,
\be\label{eq:annihilation_condition}
   \frac{\langle\sigma v\rangle} {m} \cdot \rho_{\sss NFW}(r_1)\cdot t_{\rm age}  \sim 1\,.
\ee
For $r<r_1$, the NFW profile is replaced by an isothermal one, which satisfies the Poisson equation
$v_0^2 \nabla^2\rho = -4\pi G \rho$, with one boundary condition (b.c.) being regularity at the origin~\cite{Kaplinghat:2015aga}.
Here, $v_0^2$ is the DM velocity dispersion, which is constant in the isothermal region~\footnote{There can be a transition region between the NFW and the isothermal profiles where $v_0$ varies, which for simplicity we do not attempt to capture here.}.
For the other b.c.\ of the Poisson equation, we follow Ref.~\cite{Kaplinghat:2015aga}, choosing $\rho'(r_1)$ such that the mass enclosed within $r_1$ is the same as in the original NFW profile. 
Satisfying both b.c.'s fixes the value of $v_0^2$.
Technical details about this procedure are given in Appendix \ref{app:sidm}.

As Eq.~\ref{eq:annihilation_condition} shows, the self-interaction cross section enters our results only in the  combination $\langle\sigma v\rangle \, t_{\rm age}$. 
Since $t_{\rm age}$ includes the unknown time that the binary takes to reach a $\sim 10$~pc separation through interactions with the baryonic environment, we consider it as a nuisance parameter in our analysis, normalizing  to \new{$t_{\rm age}=1$~Gyr}.
It is straightforward to rescale our results to other values of $t_{\rm age}$, which could be predicted in a more detailed analysis including the effects of stars and gas.

The SIDM halo can be matched to the spike similarly to CDM.  Ref.~\cite{Shapiro:2014oha} considered several dependences of the SI cross section on the DM relative velocity $v$,
\vskip-0.3cm
\be
    \langle\sigma_i(v)v\rangle = \sigma_0 v_0\left(v_{\rm ref}/v_0\right)^a = 
    \sigma_0 v_{\rm ref}\left(v_{\rm ref}/v_0\right)^{a-1}\,,
    \label{sig0eq}
\ee
with $a=0,1,\dots,4$.
For example, $a=4$ corresponds to Coulomb scattering, such as mediated by a massless force carrier, while $a=0$ represents isotropic scattering, resulting from a contact interaction.
\new{The choice of $v_{\rm ref}$ in 
Eq.\ (\ref{sig0eq}) is arbitrary; here we take
$v_{\rm ref} = 100\,$km/s, and quote 
values of $\sigma_0/m$, as is standard in the literature.}

Ref.~\cite{Shapiro:2014oha} identifies the spike radius  with the radius of influence of the final BH: $r_{\rm sp} = GM_\bullet/v_0^2$, so that setting $\rho_{\rm sp} = \rho_0$ in Eq.~\eqref{eq:spike_profile} gives the SIDM spike profile.
\new{The mass of the spike is typically comparable to that of the SMBH host.}
The spike density exponent depends on $a$ as $\gamma = (3+a)/4$.
We only consider values of $\sigma_0/m$ large enough so that $r_{\rm sp}\leq r_1$.
The resulting $a=0$ and $a=4$ profiles are shown in Fig.~\ref{fig:DM_spike}.
\new{For $a=0$, the isothermal core is larger and has a larger $v_0\simeq 500\,\mathrm{km}/\mathrm{s}$ than for $a=4$, for which $v_0\simeq 220\,\mathrm{km}/\mathrm{s}$.
Thus, the spike is steeper and more extended for a massless mediator.}

A more realistic cross section need not have such simple behavior.  Interactions mediated by a massive force
carrier, such as dark photons of mass $m_\gamma$, behave as $a=0$ for
$v< v_t$ and as $a=4$ for $v>v_t$, with a transition velocity $v_t\sim c\, m_\gamma/m$.
Although the DM velocity dispersion in the core is constant, it starts to rise
as \new{$v/v_0 \sim 4/11 \, (r_{\rm sp} / r)^{1/2}$} within the spike \cite{Shapiro:2014oha} and can enter the $a=4$ regime, thereby increasing $\gamma$.  We model the spike profile as Eq.\ (\ref{eq:spike_profile}) with
$\gamma = 3/4$ in the outer region and $\gamma = 7/4$ in the inner region where $v > v_t$ \new{\footnote{In the inner region, $\rho_{\rm sp}(r) = \rho_{\rm\sss NFW}(r_{\rm sp})(r_{\rm sp}/r_t)^{3/4}(r_t/r)^{7/4}$}}.
\new{These two regimes meet at the transition
radius $r_t$ at which $v=v_t$ \footnote{Assuming that $r_t < r_{\rm sp}$; otherwise one should set $r_t=r_1$ since the entire core and spike is governed by $a=4$.}.}
If $v_t < v_0$, the $a=4$ regime is applicable throughout the whole core and spike.
An example is shown in Fig.~\ref{fig:DM_spike}, where it is clear that the massive mediator interpolates between $a=0$ at large radii and $a=4$ in the inner region.

In summary, for either CDM or SIDM, the outer NFW halo parameters are determined by the BH mass $M_\bullet$, while the details of the BH spike depend on additional parameters: $\gamma$ for CDM; $\sigma_0 v_{\rm ref}/m$ and $a$ (or $v_t$ for a massive mediator) for SIDM.

\medskip
{\bf 3. SMBH merger dynamics.}
For simplicity, we assume the SMBHs to be in a circular orbit with separation $R$, angular frequency  $\omega = (GM_\bullet/R^3)^{1/2}$, and that
for separations $\lesssim 10$\,pc the orbit decays solely due to GW emission and dynamical friction with the DM.
The power in GWs is given by $P_{\rm gw} = (32/5)q^2(1+q)G^4/c^5(M_1/R)^5$~\cite{Peters:1964zz}, while the frictional power loss in the \new{common-envelope spike} is \new{(see Appendix~\ref{appC})}
\vspace{-0.1cm}
\bea
    P_{\rm df} &=&
   12\pi\, q^2\sqrt{1+q}\, 
    (G M_1)^{3/2}R^{1/2}\\
    &\times& \left[ \frac{N_1(q)}{q^{3}}\rho_{\rm sp}\left(qR\over1+q\right) + N_2(q) \rho_{\rm sp}\left(R\over 1+q\right)\right],\nn
    \label{Pdf}
\eea 
\new{where $N_{1,2}=1$ for CDM and $N_1=N_2\simeq 0.2$ for SIDM and $q=1$.}
$R(t)$ is fixed  by equating $P_{\rm df}+P_{\rm gw}$ to the rate of change of the orbital energy,
$\dot E_{\rm orb} = qGM_1^2\dot R/(2 R^2)$.  

Simple analytic solutions for $R(t)$
exist when either of the two loss terms dominate.
Since the evolution due to GW emission is well known, we focus on the dynamical friction.
Defining the characteristic timescale $t_{\rm sp} = (r_{\rm sp}^3/GM_1)^{1/2}$ and
dimensionless time and radial variables $\tau = t/t_{\rm sp}$,
\new{$x = R/(2r_{\rm sp})$}, the equation of motion takes the form
$dx/d\tau = -B x^p$, where $p = 5/2-\gamma$ and $B = f(q,\gamma)\,
\rho_{\rm sp}\, r_{\rm sp}^3/{M_1}$  with \new{$f(q,\gamma) = 96\pi\, q [(1+q)/2]^{\gamma + 1/2}(N_2 + N_1 q^{-3-\gamma})\,$}.

The timescale for hardening due to DF is then
\vspace{-0.1cm}
\begin{equation} \label{eq:t_df}
    t_{\rm df} \equiv \left.{\partial t}/{\partial\ln R} \right|_{\rm df} = ({t_{\rm sp}}/{B})\, x_{\rm crit}^{1-p},
\end{equation}
with  critical separation $x_{\rm crit}$ where DF is weakest within the DM spike.
If $\gamma\geq 3/2$, this occurs at the outer edge, $x_{\rm crit}=R_{\star}/(2r_{\rm sp})$, where $R_\star$ is the separation beyond which hardening by interactions with stars and gas is efficient; we conservatively take $R_\star=10$~pc.
For shallow spikes with $\gamma < 3/2$, DF weakens at small separations; then $x_{\rm crit}=R_{\rm gw}/(2r_{\rm sp})$, where $R_{\rm gw}$ is the separation at which GW emission becomes sufficiently strong to complete the merger. In the marginal case $\gamma=3/2$, $x_{\rm crit}^{1-p}\to\ln(R_*/R_{\rm gw})$.
For SIDM with a massive mediator, $x_{\rm crit}$ must be evaluated at the intermediate separation \new{$r_t$} at which DM particles have velocity $v_t$ \new{(see Appendix~\ref{app:sidm})}.
Using the  GW hardening timescale~\cite{Peters:1964zz}
\vspace{-0.2cm}
\begin{equation}
    t_{\rm gw} \equiv \left. \frac{\partial t}{\partial\ln R} \right|_{\rm gw} = \frac{5 c^5}{64 G^3} \frac{(2R_{\rm gw})^4}{M_1^3\,q\,(1+q)}\,,
    \label{tdfa}
\end{equation}
we find that $R_{\rm gw}=\new{0.1-0.2}\,\mathrm{pc}$ for binaries with $M_1=3\times 10^{9}\,M_{\odot}$ to merge within $0.1-1$~Gyr.
\new{We conservatively set $R_{\rm gw}=0.1$~pc for our numerical evaluations.}

\begin{figure}[t!]
\includegraphics[width=0.95 \columnwidth]
{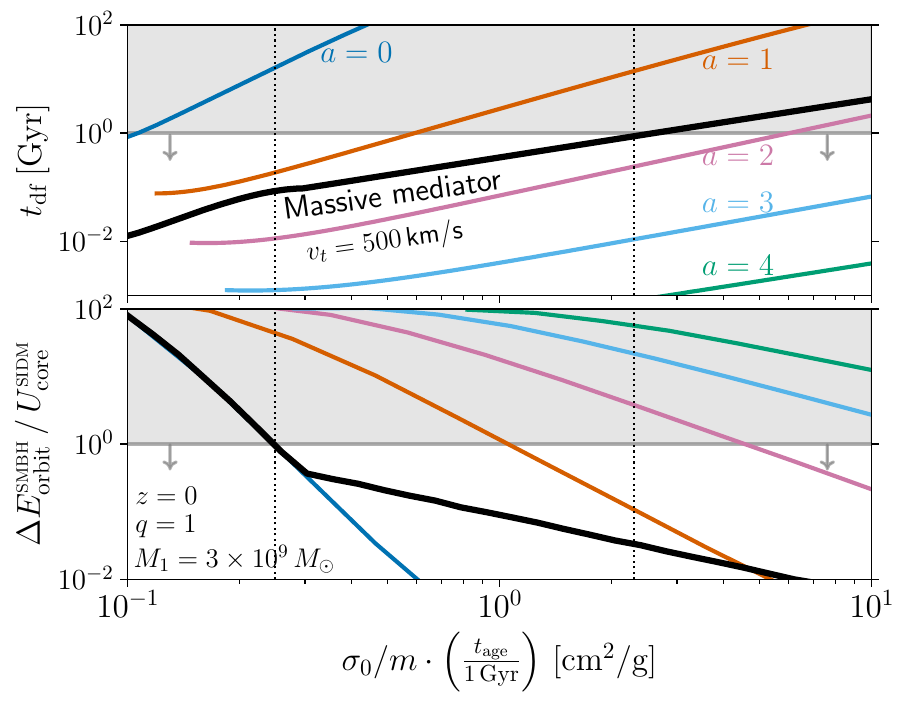}
\vspace{-0.2cm}
\caption{\new{\emph{Upper}: Time for SIDM dynamical friction to bring the SMBH separation below $0.1$\,pc where GW emission dominates, versus SI cross section.
\emph{Lower}: Ratio of orbital energy transmitted by the binary to the DM spike via dynamical friction, to the gravitational binding energy of the SIDM isothermal core.
Colors identify different velocity dependence of $\sigma(v)$, the black line corresponding to a massive mediator with $v_t=500\,\mathrm{km}/\mathrm{s}$.
Shaded regions are disfavored and dotted lines delineate the viable range for the massive mediator.
}}
\label{fig:DF_timescale}
\end{figure}

For CDM, $12\pi\rho_{\rm sp} r_{\rm sp}^3/M_1 \simeq (1+q)/2$ and thus $B=\mathcal{O}(1)$, since $q\simeq 1$ for SMBH mergers contributing most strongly to the stochastic GW background.
For binaries with $M_1\gtrsim 10^9\,M_\odot$, one finds \new{$t_{\rm sp}\simeq 2\times 10^{-3}$~Gyr. Since for a shallow spike $x_{\rm crit}\simeq 1.5\times 10^{-4}$, it follows} that $t_{\rm df}\lesssim 1$~Gyr as long as $\gamma\gtrsim 0.7$.
One would then conclude that CDM is able to solve the final parsec problem;
however, the back-reaction of the black hole motion on the spike must be taken into account \footnote{We thank J.\ Sievers for suggesting to check the energetics.}.

The energy lost by the SMBH binary when its orbit shrinks from $R_{*}$ to $R_{\rm gw}$ is 
\begin{equation}
    \Delta E_{\rm orb} = q \, G \, M_1^2 \left( R_{\rm gw}^{-1} - {R_{*}}^{-1}\right) \new{/2}.
\end{equation}
This energy heats the DM particles and should be compared to the gravitational binding energy of the spike, $U$, to estimate whether it can absorb that much energy:
\bea
    {U\over G} &=& M_\bullet \int d^{\,3}x\, {\rho_{\rm sp}(\vec x)\over|\vec x|} + 
    \int d^{\,3}x_1 d^{\,3}x_2 
        {\rho_{\rm sp}(\vec x_1) \rho_{\rm sp}(\vec x_1) \over 2|\vec x_1-\vec x_2|}\nn\\
        &=&
         4\pi M_\bullet\rho_{\rm sp}r_{\rm sp}^2 \left(
    {1-\epsilon^{2-\gamma_{\rm sp}}
    \over 2-\gamma_{\rm sp}} \right) + 4\pi^2 \rho_{\rm sp}^2 r_{\rm sp}^5 \, g(\gamma),
\eea
where $\epsilon$ is the ratio of the Schwarzschild radius to the spike radius, and $g(\gamma\lesssim 2)\sim 3$, growing to $g(7/3)\sim 20$.
Because the mass of the spike is of order $M_\bullet$,
the two contributions are comparable, and they are $4-5$ orders of magnitude smaller than $\Delta E_{\rm orb}$ for typical $M_\bullet\sim 6\times 10^{9}\,M_\odot$ contributing to the GW signal.  Hence the CDM spike is completely disrupted by the DF energy deposited.

\medskip

This obstacle can be overcome if DM has self-interactions that rethermalize and replenish the spike  sufficiently fast.
The SIDM dynamical friction timescale follows from Eq.~\eqref{eq:t_df} with  $r_{\rm sp} = G M_\bullet/v_0^2$, which depends indirectly upon $\sigma_0/m$.
\new{For our numerical evaluations, rather than the timescale we use the inspiral duration resulting from integrating the equation of motion from $R_\star$ to $R_{\rm gw}$, see Appendix~\ref{app:sidm} for more details.}
Fig.~\ref{fig:DF_timescale} (upper) shows $t_{\rm df}$ versus $\sigma_0/m$ for $a=0,1,2,3,4$ and for a massive mediator, for a representative SMBH binary \new{with the same combined mass used in Fig.~\ref{fig:DM_spike}, and thus the same host DM halo NFW parameters}.
Larger values of $a$ produce more pronounced spikes that exert more DF.
SIDM yields sub-Gyr inspiral times for values of $\sigma_0/m$ below an $a$-dependent threshold.
As $\sigma_0/m$ grows, the isothermal core becomes larger and less dense, resulting in a weaker spike that applies less DF on the SMBHs.

The energy injected in the SIDM spike by the BHs heats and disperses the DM in it, but at the same time the self-interactions repopulate the spike with DM from the isothermal core.
If this occurs fast enough, the spike survives the back-reaction and continues to harden the binary.
Since we take the spike to be in equilibrium, $t_{\rm df}$ should be larger than the SIDM core relaxation time scale $t_{\rm r} \simeq \left[{\rho_c (\sigma/m) v_0}\right]^{-1}$ (the mean time between particle collisions),
which coincides with $t_{\rm age}$ in  Eq.~\eqref{eq:annihilation_condition}.
Thus, for our approximations to be self-consistent, we demand that $t_{\rm df}=t_{\rm age}$, although this technical assumption could be relaxed in a more general approach.

If large enough, the isothermal core acts as a reservoir whose total binding energy can be sufficient to absorb the orbital energy lost by the binary. 
This puts an $a$-dependent lower limit on $\sigma_0/m$,
shown in Fig.~\ref{fig:DF_timescale} (lower).
Lower values of $a$ are favoured since they produce larger cores.
Combined with the upper limit from solving the final parsec problem, there is a range of viable $\sigma_0/m$ values that are listed in Table~\ref{tab:SIDMfpcp} in the Appendix.
The best-performing model is the massive mediator, which combines a large core in the $a=0$ regime with a steep spike in the $a=4$ phase (see Fig.~\ref{fig:DM_spike}), and prefers $\sigma_0/m \sim \mathcal{O}(\mathrm{cm}^2/\mathrm{g})$, \new{as highlighted by the dotted vertical lines in Fig.~\ref{fig:DF_timescale}}.
The preferred range of SI cross sections, which is compatible with small-scale structure constraints~\cite{Tulin:2017ara}, can be tightened by studying how GW emission from the SMBH binaries is modified by the DM.

\medskip
{\bf 4. GW spectrum.}
To calculate the total GW energy emitted by a single binary at each frequency, recall that the frequency of GWs at the source is twice the orbital frequency,
\begin{equation}\label{eq:f_to_x}
    f_s = {\omega}/{\pi} = ({2\pi t_{\rm sp}})^{-1} \sqrt{(1+q)/2} \, x^{-3/2},
\end{equation}
using the characteristic timescale and dimensionless separation \new{$x=R/(2r_{\rm sp})$} defined in the previous section.
The differential GW energy spectrum is \new{ (see Appendix \ref{appC})}
\begin{equation}
\label{spectrum_eq}
    \frac{\mathop{\mathrm{d}E_{\rm gw}}}{\mathop{\mathrm{d}f_s}} = \frac{q} {6\sqrt{2} f_s x} \new{\frac{G M_1^2}{r_{\rm sp}}} \frac{P_{\rm gw}}{P_{\rm gw} + P_{\rm df}}\,,
\end{equation}
where $x$ is understood to be a function of $f_s$ via Eq.~\eqref{eq:f_to_x}.
GW emission finishes when the two BH horizons merge at a separation $R_{\rm min} = 2 G M_1 (1+q)$, resulting in a sharp cutoff of the spectrum at high frequency. The GW signal produced by a population of cosmological SMHB mergers is described by the characteristic strain~\cite{Phinney:2001di,Chen:2016zyo}
\begin{equation}
    h_c^2(f) = \frac{4G}{\pi c^2 f}\int\mathop{\diff z}\mathop{\diff M}\mathop{\diff q} \frac{\mathop{\diff^3 n}}{\mathop{\diff z}\mathop{\diff M}\mathop{\diff q}} 
    \frac{\mathop{\diff E_{\rm gw}}}{\mathop{\diff f_s}}\,,
    \label{hceq}
\end{equation}
where $f = f_s/(1+z)$ is the frequency of the GW at the detection point.

The NANOGrav analysis \cite{NANOGrav:2023hfp} includes a phenomenological parameter for the total hardening timescale. Its
posterior distribution is peaked at the lower bound of the range considered, $0.1$~Gyr, signaling a preference for fast coalescence, with a $1$-$\sigma$ region extending to a few Gyr. We therefore approximate the inspiral as being instantaneous compared to the Hubble time, and explore values of $t_{\rm age}$ in the $10$~Myr\,-\,$1$~Gyr range.
This is consistent with Section 3, which showed that DF from the DM spike can yield a sub-Gyr merger time for typical SMBHs from separations as large as 10~pc, beyond which interactions with stars and gas are assumed to be effective.
We neglect possible effects of ambient stars or gas on the GW waveform within the PTA frequency range, to clearly illustrate the effects of DM.

The parametrization of $\diff^{3}n$, described in Appendix~\ref{app:Nanograv_param}, is based on Ref.~\cite{Chen:2018znx}, used by NANOGrav \cite{NANOGrav:2023hfp} and EPTA~\cite{EPTA:2023xxk}.
We expect the astrophysical parameters determining it to be largely unchanged by DM effects, except for the overall normalization of the signal that is sensitive to the hardening timescale. We therefore fix them to the best-fit values found by Ref.~\cite{NANOGrav:2023hfp} and allow only the normalization parameter $\psi_0$ to vary.

\begin{figure}[t!]
\!\!\!\!\!\!\!\!
\includegraphics[width=0.99 \columnwidth]{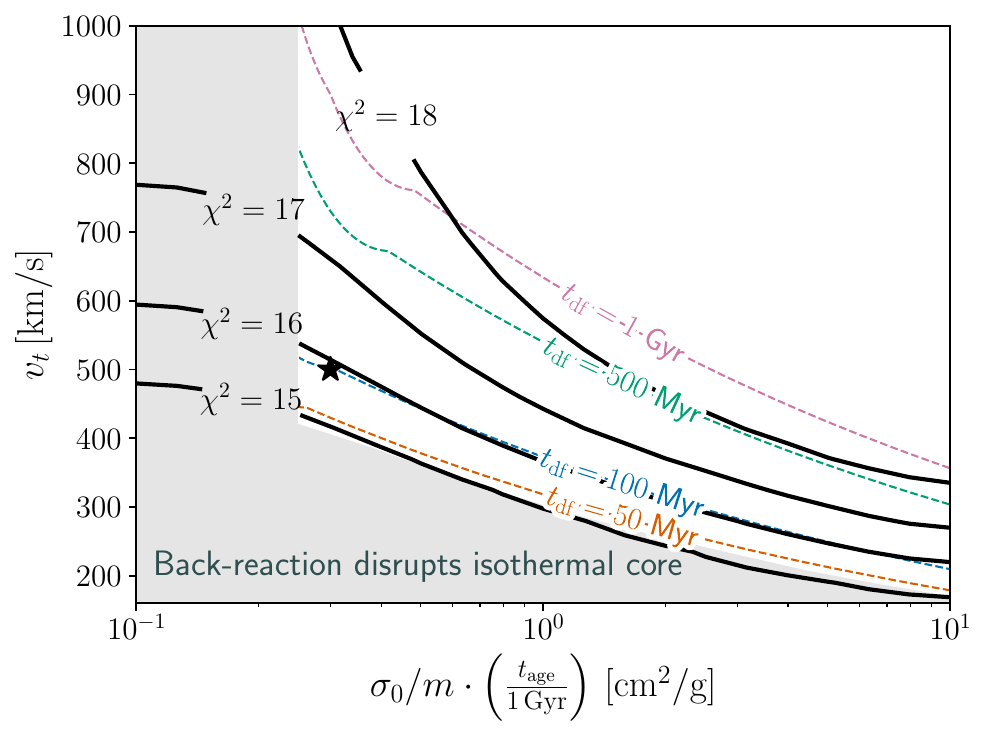}
\vspace{-0.2cm}
\caption{Contours of $\chi^2$ from the fit to the characteristic strain spectra of NANOGrav~\cite{NANOGrav:2023hfp}, PPTA~\cite{Reardon:2023gzh}, and EPTA~\cite{EPTA:2023xxk}, for a SIDM model with a massive mediator, in the plane of $v_t$ (velocity at which the SI transitions from contact-like to long-range) versus interaction cross section per DM mass at $100\,\mathrm{km}/\mathrm{s}$.
In the grey shaded region, the orbital energy lost by SMBHs with equal $3\times 10^9\,M_\odot$ masses at $z=0$ is larger than the gravitational binding energy of the SIDM core.
Dashed lines show contours of the DF timescale, assuming $t_{\rm age}=t_{\rm df}$.
}
\label{fig:contours}
\end{figure}

\begin{figure}[t!]
\!\!\!\!\!\!\!\!
\includegraphics[width=0.99 \columnwidth]{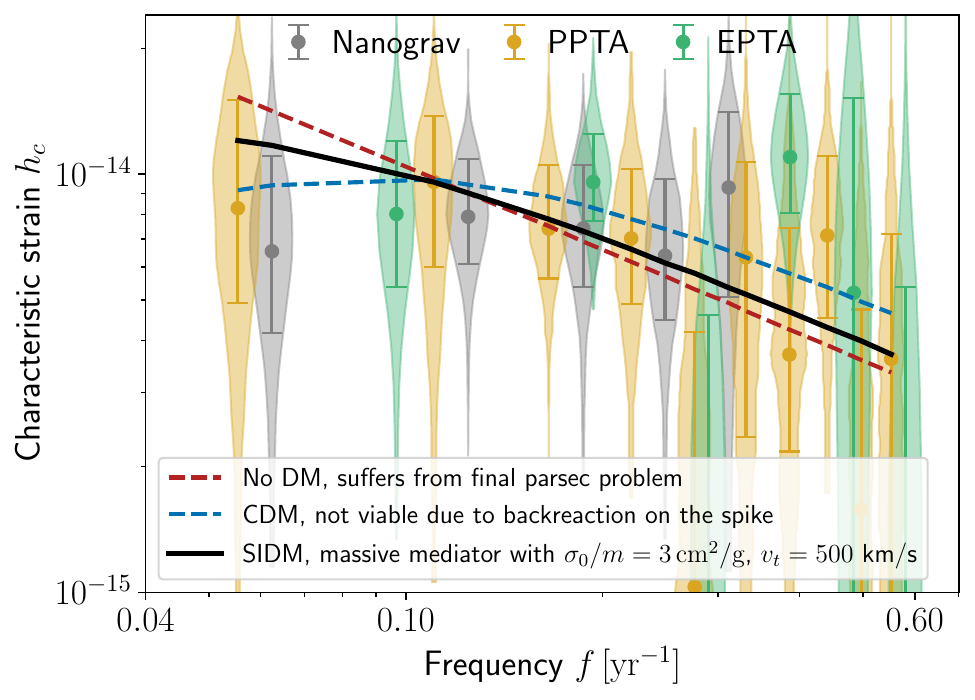}
\vspace{-0.2cm}
\caption{Selected characteristic strain spectra compared to PTA data. The SIDM model corresponds to the black star in Fig.~\ref{fig:contours}, for which $t_{\rm df}=100$~Myr.}
\label{fig:hc}
\end{figure}

We extract central $h_c$ values and upper and lower error bars for the first five NANOGrav~\cite{NANOGrav:2023gor}, ten PPTA~\cite{Reardon:2023gzh}, and six EPTA~\cite{EPTA:2023xxk} frequency bins,
by fitting two one-sided gaussians to the probability distributions represented by the violin plots;
see Table~\ref{tab:data_points} in the Appendix.  From these, 
$\chi^2$ values are calculated for model predictions using Eq.~\eqref{hceq}. 
The characteristic strain spectrum for the best-fitting model in each category is shown in Fig.~\ref{fig:hc}. 
The best fits among them give $\chi^2 \lesssim 15$, which is lower than the expected $\sim 21$ due to correlations between the different frequency bins in the data~\footnote{Taking into account the correlations between different energy bins and/or experiments in our statistical treatment is beyond the scope of this work.}.
The normalization is treated as a nuisance parameter, and for our best-fitting models is $\psi_0 \sim -2.5$. 
This is significantly smaller than that found in the fiducial NANOGrav analysis~\cite{NANOGrav:2023hfp} and better matches astrophysical expectations~\cite{2016ApJ...830...83C,Chen:2018znx}, but the conclusion may vary when the finite duration of the mergers is taken into account.

As is visible in Fig.~\ref{fig:hc}, the presence of DM improves the fit by softening the gravitational wave spectrum at low frequencies, where energy is being lost to DF rather than emitted in GWs.
Although a CDM spike with $\gamma\simeq1.5$ appears to give a good fit, the destructive back-reaction undermines this result.
SIDM with a single power-law velocity dependent cross section can only produce a moderate softening due to the requirement of having a large enough isothermal core to survive back-reaction. The minimum $\chi^2$ value within the viable $\sigma_0/m$ range can be found
in Table~\ref{tab:SIDMfpcp} in the Appendix for these models.

SIDM with a massive mediator is the only model among the ones studied that can simultaneously absorb the DF heat and produce a \new{noticeable} softening in the GW spectrum.
As shown in Fig.~\ref{fig:contours} \new{as a function of the two free parameters of the model}, we find good fits to the PTA data, compatible with merging well within 0.1 Gyr, for $v_t\sim 300$\,-\,$600\,$km/s and $\sigma_0/m \sim 2.5$\,-\,$25$ cm$^2/\mathrm{g}\cdot(100\,\mathrm{Myr}/t_{\rm age})$. Recall that we impose $t_{\rm age}=t_{\rm df}$ as given by the colored dashed contours in Fig.~\ref{fig:contours}.

\medskip
{\bf 5. Conclusions.} Despite astrophysical uncertainties about their detailed nature, there is no doubt that dark matter spikes
exist around supermassive black hole binaries, and thus contribute to the dynamical friction accelerating the decay of their orbit. 
We have shown that well-motivated models of velocity-dependent self-interacting dark matter 
have two correlated and desirable effects: robustly resolving the long-standing final parsec problem, 
thus allowing GW emission to finish the inspiral process, and softening the GW strain spectrum at low frequencies, matching the feature hinted at by PTA data.
In contrast, collisionless CDM spikes are incapable of absorbing the frictional heat and are destroyed
by the merger.

It is encouraging that the properties of the self-interaction cross section favored by the PTA signal
is compatible with what was already proposed to solve the small-scale structure problems of CDM. 
In particular, the required magnitude and broken power law velocity dependence can 
reconcile the different cross section values needed to explain observations of galaxy and galactic cluster core sizes~\cite{Kaplinghat:2015aga}.

\new{Our preliminary study opens the way to using gravitational wave signals from supermassive black hole mergers as a probe of dark matter microphysics.
Ways to sharpen the predictions include accounting for the finite duration of the inspiral in an improved statistical analysis, using a more realistic velocity dependence for the SIDM scattering cross section, and performing numerical simulations to validate the analytical calculations for the back-reaction on the DM profile.}

\medskip
{\bf Acknowledgments.}  We thank 
A.\ Benson, L.\ Combi, D.\ Curtin, C.\ Mingarelli, M.\ Reina-Campos, J.\ Sievers, S.\ Tulin,  and G.-W.\ Yuan for helpful discussions.  This work was supported by the Natural Sciences and Engineering Research Council (NSERC) of Canada.

\begin{appendix}
\label{sec:appendix}
\section{SMBH and halo relations}
\label{app:halo}

\begin{figure}[t]
\centerline{\!\!\!\!\!\!\!\!\!\!
\includegraphics[width=0.99 \columnwidth]{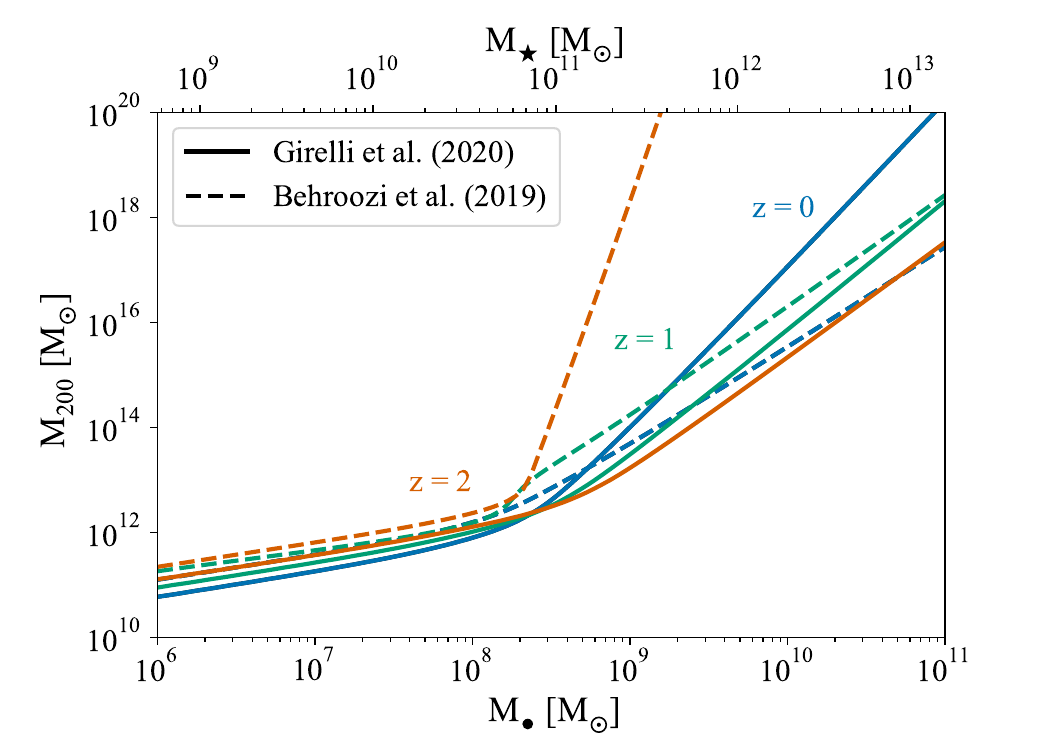}}
\vspace{-0.2cm}
\caption{The halo-to-black hole mass relation $M_{200}-M_{\bullet}$ for several values of redshift z. The solid lines use the stellar-to-halo mass relation from \cite{Girelli:2020goz}, while the dashed lines use the relation from \cite{Behroozi:2019kql}. The upper x-axis shows the resulting stellar mass $M_{\star}$ from combining equations \ref{eq:M_to_Mbulge} and \ref{eq:Mbulge_to_Mstar}.}
\label{fig:BHtoHaloMass}
\end{figure}

The effect of the DM spike is correlated with $M_\bullet$ by a series of relations involving the (pseudo)bulge mass $M_{\rm bul}$ of the host galaxy and its stellar mass $M_{\star}$. 
At a given redshift $z$, the halo-to-stellar mass relation is a function $[M_{200}/M_{\star}](z)$,
which for our fiducial results we take from Ref.~\cite{Girelli:2020goz}:
\be\label{eq:M200_to_Mstar}
    {M_{200}\over M_{\star}}(z) = {1\over  2A(z)}\left[\left(\frac{M_{200}}{M_{A}(z)}\right)^{-\beta(z)}\!\!\!\! +\ \left(\frac{M_{200}}{M_{A}(z)}\right)^{\gamma(z)}\right]\,,   
\ee
where polynomial fits to the functions $A,\,M_A,\,\alpha,
\,\beta,\,\gamma$ of $z$ are given.
The galaxies of interest for the PTA signal have $M_\star \gtrsim 10^{11}\,M_\odot$, at the heavy end of the distribution where uncertainties are large.
We have confirmed that our results are robust against using a different stellar-to-halo mass relation, in particular that of Ref.~\cite{Behroozi:2019kql}, for which
\be\label{eq:M200_to_Mstar_2}
 {M_{200}\over M_{\star}} = {(M_1/M_{200})^\alpha + (M_1/M_{200})^\beta\over 10^{\epsilon +\gamma\exp[-(x/\delta)^2/2]} }\,,
\ee
where $\log M_1$, $\alpha$, $\beta$, $\gamma$ are linear functions of $z$ and
$x = \log_{10}(M_{200}/M_1)$.
As can be seen in Fig.~\ref{fig:BHtoHaloMass}, the latter relation predicts significantly smaller DM haloes at the heavy end of the spectrum at low redshift.

To relate the SMBH mass to the stellar mass of the galaxy, we use the black-hole-to-bulge mass relation from Ref.~\cite{Kormendy:2013dxa},
\begin{equation}\label{eq:M_to_Mbulge}
    \log_{10}\left( \frac{M_\bullet}{M_\odot} \right) = 8.7 + 1.1 \log_{10} \left( \frac{M_{\rm bulge}}{10^{11}M_\odot} \right).
\end{equation}
This phenomenological fit has a normally distributed scatter of $\sim 0.3$ dex, which we do not try to simulate here.
The bulge mass is related to the stellar mass by~\cite{Chen:2018znx}
\begin{equation}\label{eq:Mbulge_to_Mstar}
    M_{\rm bulge} = f_{\star,\mathrm{bulge}} M_\star,
\end{equation}
where $f_{\star,\mathrm{bulge}} = 0.615 + df_{\star}$, with
\begin{equation}
    df_{\star} = \begin{cases}
        0,\qquad\qquad\qquad\qquad\ \! \quad\mathrm{if}\, M_\star \leq 10^{10}M_\odot\\
        \frac{\sqrt{6.9}\,\mathrm{exp}\left( \frac{-3.45}{\log_{10} M_\star - 10} \right)}{(\log_{10} M_\star - 10)^{1.5}}\,,\quad\mathrm{if}\, M_\star > 10^{10}M_\odot
    \end{cases}
\end{equation}
The small correction $df_\star$ peaks around $M_*\sim 10^{12}\,M_\odot$, corresponding to $M_\bullet\sim\mathrm{few}\times 10^9\,M_\odot$. 

By numerically inverting the above relations, one can find $M_{200}$ for a given $M_\bullet$ and redshift $z$ (see Fig.~\ref{fig:BHtoHaloMass}).
This provides one input for fixing the outer part of the DM halo density profile, which we take to be NFW~\cite{Navarro:1995iw},
\be
    \rho_{\sss NFW}(r) = \frac{\rho_s}{\left( {r/ r_s}\right)\left(1 + {r/r_s}\right)^{2}}\,.
\ee
The virial radius $R_{200}$ is defined to be that which contains the mass $M_{200} = 4\pi\int dr\,r^2\rho(r)$, which corresponds to 200 times the mass given by 
the volume $4\pi R_{200}^3/3$ times the critical density $\rho_c(z)$.
The NFW parameters $\rho_s(z)$ and $r_s(z)$ are 
constrained by the concentration parameter $c_{200}(z) = R_{200}/r_s$, which evolves with $z$ as given in Ref.~\cite{Klypin:2014kpa}:
\begin{equation}
    c_{200}(z) ={ C_{c}(z)\over\left(M_{200}/M_{\rm ref}\right)^{\gamma_{c}(z)}}\left[1+\left(\frac{M_{200}}{M_{c}(z)}\right)^{0.4}\right]\,,
\end{equation}
with $M_{\rm ref} = 10^{12}h^{-1}M_{\odot}$, and the
functions $C_c,\, \gamma_c$ and $M_c$ are tabulated
in Ref.\ \cite{Klypin:2014kpa}.
These relations numerically determine  the NFW parameters $\rho_s$ and $r_s$ from $M_\bullet$ and $z$. 

\section{SIDM core profile \new{and merger time}}
\label{app:sidm}
\begin{figure}[t]
\centerline{
\includegraphics[scale=0.5]{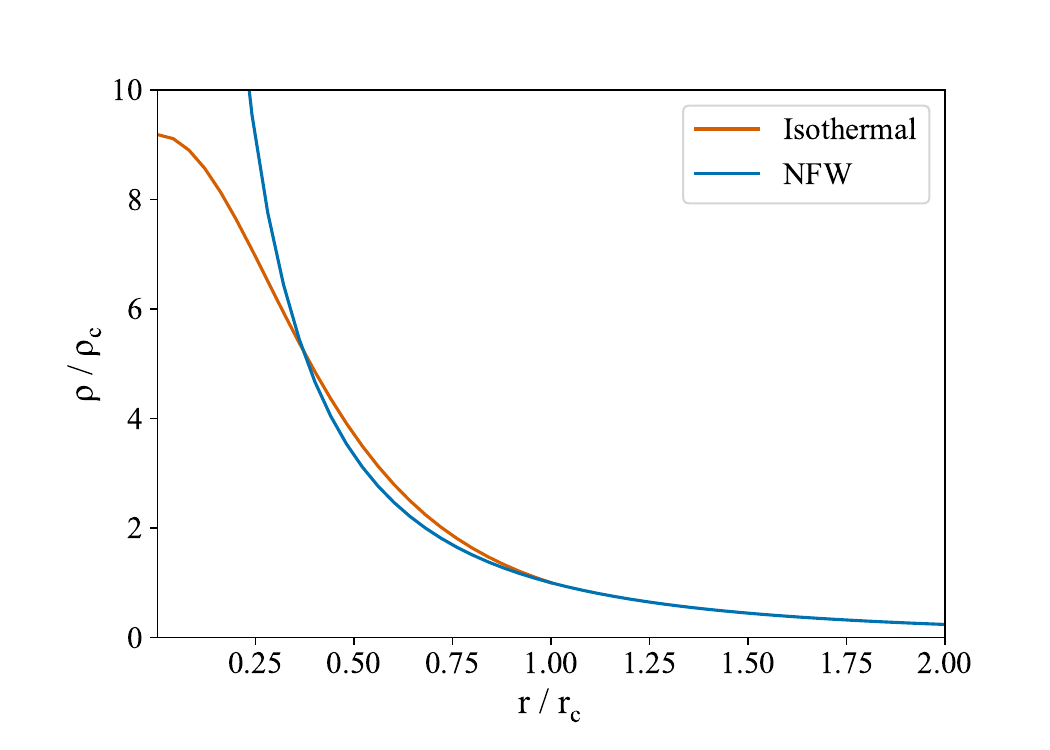}
}
\centerline{
\includegraphics[scale=0.5]{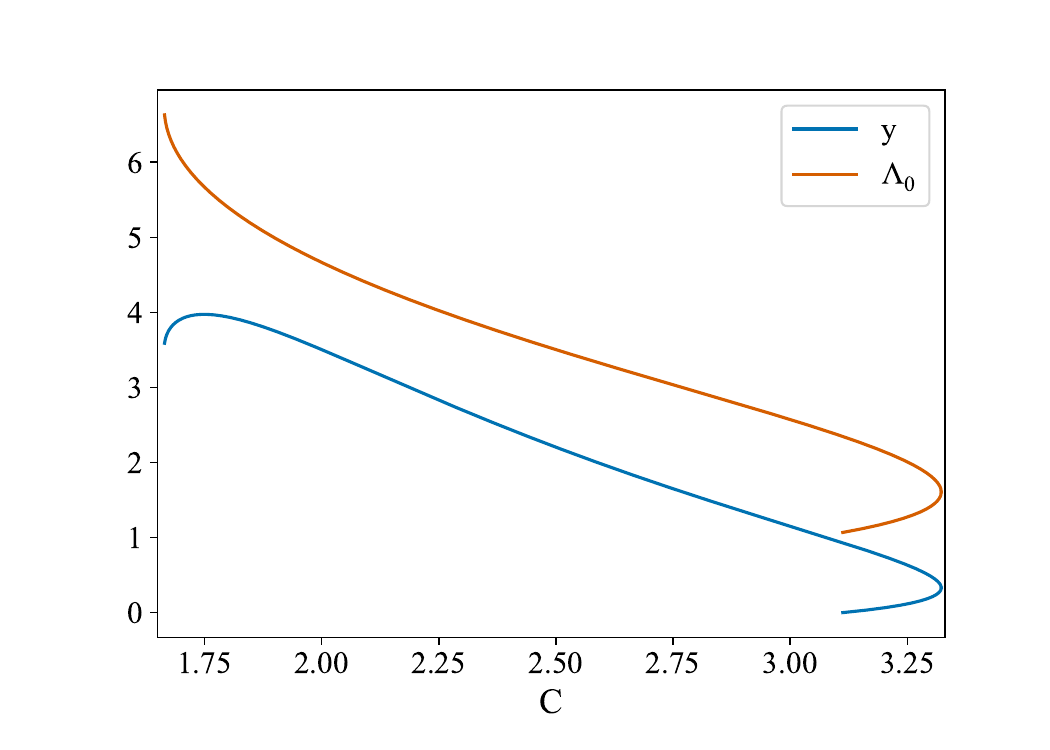}
}
\vspace{-0.2cm}
\caption{Top: example of cored density profile (orange) versus original NFW
profile (blue), for $C=3.17$ and $y=0.815$.  Bottom:
$\Lambda_0$ (orange) and $y=r_1/r_s$ (blue) as a function of $C$ for 
cored halos satisfying the Poisson equation.
}
\label{fig:C}
\end{figure}

To construct the isothermal core region of the SIDM halo profile, one must solve the 
Poisson equation 
\begin{equation}
    v_{0}^{2} \nabla^{2} \ln \rho = -4 \pi G \rho
        \label{eq:poisson}
\end{equation}
between $r=0$ and $r=r_1$ with boundary conditions
$\rho'(0)=0$ and $\rho(r_1)\equiv \rho_c = \rho_{\sss NFW}(r_1)$, subject to the constraint that the mass enclosed within $r_1$ is conserved
relative to its value in the original NFW profile.
Using the dimensionless variables $w=r/r_1$ and $\Lambda = \ln(\rho/\rho_c)$, Eq.\ (\ref{eq:poisson}) becomes
\be
    \Lambda'' + {2\over w}\Lambda' = -C e^\Lambda\,,
\ee
where $C = 4\pi G \rho_c r_1^2/v_0^2$.  The boundary conditions become $\Lambda'(0)=0$ and $\Lambda(1)=0$.  For a given value of $C$, this can be solved by shooting from $w=0$: take $\Lambda'(0)=0$ and $\Lambda(0)=\Lambda_0$, and vary $\Lambda_0$ until $\Lambda(1) = 0$.  
Next one must satisfy the mass conservation condition:
\be
   { \int_0^1 w^2 e^\Lambda dw
  } =  \int_0^1 {w (1+y)^2\over (1+wy)^2}\, dw\,,
\ee
where $y=r_1/r_s$ ($r_s$ is the scale radius of the NFW profile.)
This requires finding the right value of $C$,
which amounts to determining $v_0$ since $\rho_c$ and $r_1$ are fixed by Eq.\ (\ref{eq:annihilation_condition}).  An example of the resulting profile is shown in
Fig.~\ref{fig:C} (top).

A subtlety in this procedure is that there can be more than one possible value of $\Lambda_0$  satisfying the boundary conditions for a given value of $y$.
For example, with $y=0.815$
 there are two possible solutions, and it is necessary to choose the larger one in order to satisfy the mass conservation requirement.  The required values are $\Lambda_0 = 2.22$ and $C = 3.17$.
This gives rise to a small discontinuity in $\rho'$ at $w=1$, from $-1.9$ to $-2.3$, which is 
hardly noticeable in 
Fig.\ \ref{fig:C} (top panel). 

In practice, instead of shooting, it is numerically efficient to fix $C$ and solve for
$y(C)$, and $\Lambda_0(C)$, and then numerically invert the relations.  By doing so, one can avoid having to re-solve the Poisson equation for each different set of
NFW halo parameters and self-interaction
cross section.  The resulting functions are shown in Fig.\ \ref{fig:C} (bottom).  Then, a given $M_\bullet$ determines the NFW parameters, while $\sigma_0/m$ which determines $r_1$ hence $y$, and the velocity dispersion in the isothermal core is given by
\be
    v_0^2 = 4\pi {y\over (1+y)^2C(y)} G \rho_s r_s^2 \, .
\label{v02eq}
\ee

\new{For the massive mediator model with velocity dependent cross section
\be
	\sigma(v) = {\sigma_0\over 1 + (v/v_t)^4 }\cong
\sigma_0\left\{\begin{array}{ll}1,& v < v_t\\
(v_t/v)^4,& v > v_t\end{array}\right.\,, 
\ee
one must verify that the solution within the core is self-consistent.
We first compute $v_0$ from Eq.\ (\ref{v02eq}) assuming $a=0$
(i.e., $\sigma=\sigma_0$) in Eq.~(\ref{eq:annihilation_condition}), which determines $r_1$.  If $v_0>v_t$, then $r_1$ and the other core parameters must be
recomputed for $a=4$ (i.e., $\sigma = \sigma_0 (v_t/v)^4$).

\new{As stated in the main text, we assume the velocity dispersion of the SIDM particles to be constant in the core and to vary as
\begin{equation}
    \frac{v(r)}{v_0} = \frac{7}{11} + \frac{4}{11}\left( \frac{r_{\rm sp}}{r} \right)^{1/2}
    \label{vreq}
\end{equation}
within the spike.
This slightly deviates from the analytic approximation proposed in~\cite{Shapiro:2014oha} to ensure continuity of the velocity dispersion profile at $r=r_{\rm sp}$.
Thus, if $v_0 < v_t$, the $a=0$ to $a=4$ transition occurs in the spike, at a transition radius $r_t$ such that $v(r_t)=v_t$ in Eq.\ (\ref{vreq}).
}

To compute $t_{\rm df}$ in the dark photon model, three regimes must be considered. 
If the transition occurs at $r_t< r_{\rm gw}$, the entire spike is in the $a=0$ region, 
and the equation of motion for the dimensionless separation $x = R/(2r_{\rm sp})$ is $dx/d\tau = - B x^{7/4}$ with $B = 192\pi N_i \rho_{\rm sp} r_{\rm sp}^3/M_1$ for
the case of $q=1$ (see above Eq.\ (\ref{eq:t_df})).
The solution is 
\be
    t_{\rm df} = {4 t_{\rm sp}\over 3 B}
        \left(x_{\rm gw}^{-3/4} - x_*^{-3/4} \right).
\ee
If the transition region occurs outside of $R_*$, then the $a=4$ solution applies everywhere, and the equation of motion is $dx/d\tau = - 2B x_t x^{3/4}$. The
time required to shrink from $R_*$ to $R_{\rm gw}$ is 
\be
    t_{\rm df} = 2 {t_{\rm sp}\over B x_t}\left(x_*^{1/4} - x_{\rm gw}^{1/4}\right).
\ee
If the transition region is inside the spike, the equation of motion is $dx/d\tau = -B x^{7/4}$
in the outer region $x>x_t$, and by continuity 
$dx/d\tau = -2B x_t x^{3/4}$ in the inner region $x< x_t$.
The resulting time interval is 
\be
    t_{\rm df} = 2 {t_{\rm sp}\over B}\left(\sfrac53 x_t^{-3/4} - {x_{\rm gw}^{1/4}\over x_t} -\sfrac23 x_*^{-3/4}\right).
    \label{xtinspike}
\ee
The expression (\ref{xtinspike}) joins continuously with the previous ones if $x_t = x_{\rm gw}$ or $x_*$.
}

\section{Derivation of DF power loss formula and single-merger GW spectrum}
\label{appC}
\new{
To derive the dynamical friction power loss Eq.~\eqref{Pdf}, we start from the classical expression for the DF force originally derived by Chandrasekhar~\cite{Chandrasekhar:1943ys} (see, e.g.~\cite{Gualandris:2007nm})
\begin{equation}\label{eq:Fdf}
    F_{\rm df} = -4\pi G^2 \rho_{\rm sp} M_{\rm bh}^2 \log\Lambda \frac{N(<v_{\rm bh})}{v_{\rm bh}^2}.
\end{equation}
Here, $M_{\rm bh}$ and $v_{\rm bh}$ are the black hole mass and velocity, $\log\Lambda$ is the Coulomb logarithm which we take to be $3$~\cite{Gualandris:2007nm}, and $N$ denotes the fraction of DM particles that have velocities $<v_{\rm bh}$.
As a function of the binary separation $R$, $v_{\rm bh1} = q(GM_1/(1+q) R)^{1/2}$ and $v_{\rm bh2} = (GM_1/(1+q) R)^{1/2}$, for the heavier ($M_{\rm bh}=M_1$) and lighter ($M_{\rm bh}=qM_1$) BH, respectively.

For CDM $N\simeq 1$, but for SIDM the velocity dispersion in the inner core is larger due to the self-interaction driven thermalization and this fraction can be smaller.
Assuming a Maxwellian distribution for the SIDM particles, one has
\begin{equation}
    N_i = \mathrm{erf}\left(\frac{u_i}{\sqrt{2}}\right) - \sqrt{\frac{2}{\pi}} \, u_i\,\mathrm{e}^{-u_i^2/2},
\end{equation}
$i=1,2$. Here, $u_i$ is the ratio between $v_{\rm bh,i}$ and the DM velocity dispersion for the $i$th BH,
\begin{equation}
    u_i = \begin{cases}
        \frac{11}{4}\, q^{3/2}\,(1+q)^{-3/2}, &\quad i=1  \\
        \frac{11}{4}\, (1+q)^{-3/2}, &\quad i=2
    \end{cases}
\end{equation}
which stays constant throughout the spike since both velocities scale as $r^{-1/2}$.
Substituting these expressions into Eq.~\eqref{eq:Fdf}, together with the radial coordinate of each BH from the center of mass, $r_1=qR/(1+q)$ for the heavier BH and $r_2 = R/(1+q)$ for the lighter BH, one arrives at the expression for $P_{\rm df} = F_{\rm df}\,v_{\rm bh}$ given in Eq.~\eqref{Pdf}.
}

The spectrum of gravitational waves, Eq.\ (\ref{spectrum_eq}) follows from the power emitted in GWs, defined above Eq.\ (\ref{Pdf}), and the relation between frequency and orbital separation,
Eq.\ (\ref{eq:f_to_x}), by using the chain rule:
\be
    {dE_{\rm gw}\over df}(f) = {P_{\rm gw}\over df/dt}
    = {P_{\rm gw}\over \sfrac32 f\dot x/x}
    = {G M_1 M_2\over 6\sqrt{2} f x\, r_{\rm sp} }{P_{\rm gw}\over P_{\rm tot}}\,.
\ee
To obtain the last equality, we use $\dot E = -\sfrac{1}{4\sqrt{2}}(G M_1 M_2/r_{\rm sp})(\dot x/x^2) = - P_{\rm tot}$, the total rate of energy loss from the circular orbit.  We assume that GW emission and dynamical friction are the only significant sources of energy loss, so that $P_{\rm tot} = P_{\rm gw} + P_{\rm df}$. Alternatively, one can derive Eq.\ (\ref{spectrum_eq}) from the standard expression for the GW spectrum from a merging binary \cite{Phinney:2001di}, as in Eq.\ (16) of Ref.~\cite{Shen:2023pan}, by averaging over the inclination angle of the orbit; we have confirmed that the two methods lead to the same result.

\section{Parametrization of the SMBH binary population}\label{app:Nanograv_param}

\new{The density of SMBH mergers as a function of redshift is related to the differential galaxy merger rate,
\begin{equation}
    \frac{\mathop{\diff^3 n}}{\mathop{\diff z}\mathop{\diff M}\mathop{\diff q}} = \frac{\mathop{\diff^3 n_{\rm g}}}{\mathop{\diff z}\mathop{\diff M_\star}\mathop{\diff q_\star}} \frac{\mathop{\diff M_\star}}{\mathop{\diff M}} \frac{\mathop{\diff q_\star}}{\mathop{\diff q}}\,,
\end{equation}
where $M_\star$ is the stellar mass of the primary merging galaxy and $q_\star$ is the mass ratio of the two galaxies involved.}
We parametrize the differential galactic merger rate as~\cite{Chen:2018znx,NANOGrav:2023hfp}
\begin{equation}
    \frac{\mathop{\diff^3 n_{\rm g}}}{\mathop{\diff z}\mathop{\diff M_\star}\mathop{\diff q_\star}} = \frac{\Psi(M_\star,z')}{M_\star} \frac{P(M_\star,q_\star,z')}{T_{\rm g\hbox{-}g}(M_\star,q_\star,z')} \frac{\mathop{\diff t}}{\diff z'},
\end{equation}
in terms of the galaxy stellar-mass function $\Psi$, the galaxy pair fraction $P$, and the galaxy merger time $T_{\mathrm{g}\hbox{-}\mathrm{g}}$.
Because the galaxy merger can take a significant time, the quantities on the right-hand side must be evaluated at the advanced redshift $z'$, such that $t(z)-t(z')=T_{\mathrm{g}\hbox{-}\mathrm{g}}(z')$.
We use
\be
    \frac{\mathop{\diff t}}{\diff z} = \frac{1}{(1+z)H(z)}\,,
\ee
where
\be
    H(z) = H_0\left[\Omega_\Lambda + (1+z)^3\Omega_m\right]^{1/2}
\ee
and $H_0=67.4\,\mathrm{km}\,\mathrm{s}^{-1}\mathrm{Mpc}^{-1}$, $\Omega_m=0.315$, and $\Omega_\Lambda=0.685$~\cite{Planck:2018vyg}.
This, together with $t(z=0)=13.79$~Gyr~\cite{Planck:2018vyg}, allows to calculate $z(t)$.
The galactic stellar mass function follows
\begin{equation}
    \Psi(M_\star,z) = \Psi_0 \left( \frac{M_\star}{M_\Psi} \right)^{\alpha_\Psi} \mathrm{exp}\left( -\frac{M_\star}{M_\Psi} \right),
\end{equation}
with
\begin{align}
    \log_{10}\left( \frac{\Psi_0}{\mathrm{Mpc}^{-3}} \right) &= \psi_0 + \psi_z \cdot z,\\
    \log_{10}\left( \frac{M_\Psi}{M_\odot} \right) &= m_{\psi0} + m_{\psi z} \cdot z,\\
    \alpha_\Psi &= 1 + \alpha_{\psi 0} + \alpha_{\psi z}\cdot z.
\end{align}
As detailed in Table~\ref{tab:SMBH_pop_params}, all of this parameters are kept fixed except for $\psi_0$, which is varied to adjust the total normalization of the differential merger rate.
The galaxy pair fraction is described by
\begin{equation}
    P(M_\star,q_\star,z) = P_0 \, (1+z)^{\beta_{p0}},
\end{equation}
while the galaxy merger time is given by
\begin{equation}
    T_{\mathrm{g}\hbox{-}\mathrm{g}}(M_\star,q_\star,z) = T_0\, (1+z)^{\beta_{t0}} \,q_\star^{\gamma_{t0}}.
\end{equation}
All the parameters in these last two functions are fixed to the values shown in Table~\ref{tab:SMBH_pop_params}.

To relate the SMBH mass to the stellar mass of the galaxy, we use the relations described in Appendix~\ref{app:halo}.
Note that in our analysis we do not include any scatter in Eq.~\eqref{eq:M_to_Mbulge}.
We do not expect this simplification to affect our conclusions since the NANOGrav posterior distribution for the variance in the relation is peaked at zero~\cite{NANOGrav:2023hfp}.

\setlength{\tabcolsep}{12pt}
\begin{table}[t!]
\begin{tabular} { c  c | c  c }
Parameter & Value & Parameter & Value \\
\hline
    $\psi_0$ & Free & $P_0$ & $0.033$ \\
    $\psi_z$& -0.6 & $\beta_{p0}$ & $1$ \\
    $m_{\psi0}$ & $11.5$ & $T_0$ & $0.5$~Gyr \\
    $m_{\psi z}$ & $0.11$ & $\beta_{t0}$ & $-0.5$ \\
    $\alpha_{\psi0}$ & $-1.21$ & $\gamma_{t0}$ & $-1$ \\
    $\alpha_{\psi z}$ & $-0.03$
\end{tabular}
\caption{List of parameters describing the SMBH merger population along with the values used in our numerical evaluations, based on the fiducial analysis of~\cite{NANOGrav:2023hfp}.}
\label{tab:SMBH_pop_params}
\end{table}

\setlength{\tabcolsep}{4pt}
\begin{table}[t!]
\begin{tabular} { c  c  c }
$a$ &  Viable cross section $\left[\mathrm{cm}^2/\mathrm{g}\right]$ &  $\chi^2_{\rm min}$ \\
\hline
0 & None & -- \\
1 & None & -- \\
2 & \phantom{}$4.5 \lesssim \sigma_0/m \cdot \left( \frac{t_{\rm age}}{1\,\mathrm{Gyr}} \right)\lesssim 6$ & 19.2 \\
3 & \phantom{aa}$20 \lesssim \sigma_0/m \cdot \left( \frac{t_{\rm age}}{1\,\mathrm{Gyr}} \right)\lesssim 90$ & 18.8 \\
4 & \phantom{aaa}$ 50\lesssim \sigma_0/m \cdot \left( \frac{t_{\rm age}}{1\,\mathrm{Gyr}} \right) \lesssim 1600$ & 17.6 
\end{tabular}
\caption{Viable values of the dark matter self-interaction cross section with velocity-dependence parametrized by Eq.~\eqref{sig0eq}. 
The lower limit ensures that the isothermal core is large enough to absorb the frictional energy, while the upper limit comes from requiring $\tau_{\rm df}<1$~Gyr. The last column shows the test statistic of the PTA $h_c$ fit at the best-fit point within each range of cross sections. For reference, the GW-only fit without DM dynamical friction has $\chi^2 = 19.3$.}
\label{tab:SIDMfpcp}
\end{table}

\setlength{\tabcolsep}{10pt}
\begin{table}[t!]
\begin{tabular} { c c | c c c c }
$f\,[\mathrm{yr}^{-1}]$ & $h_c/10^{-15}$ & $f\,[\mathrm{yr}^{-1}]$ & $h_c/10^{-15}$ \\
\hline
\multicolumn{2}{c |}{NANOGrav~\cite{NANOGrav:2023gor}} & \multicolumn{2}{c}{PPTA~\cite{Reardon:2023gzh}} \\
    0.062 & 6.5{\raisebox{0.5ex}{\tiny$\substack{+4.5 \\ -2.4}$}} & 0.055 & 8.3{\raisebox{0.5ex}{\tiny$\substack{+6.8 \\ -3.4}$}} \\
    
    0.12 & 7.9{\raisebox{0.5ex}{\tiny$\substack{+3.0 \\ -1.8}$}} & 0.11 & 9.6{\raisebox{0.5ex}{\tiny$\substack{+4.2 \\ -3.6}$}} \\
    
    0.19 & 7.4{\raisebox{0.5ex}{\tiny$\substack{+3.1 \\ -2.0}$}} & 0.17 & 7.4{\raisebox{0.5ex}{\tiny$\substack{+3.1 \\ -1.8}$}} \\

    0.25 & 6.4{\raisebox{0.5ex}{\tiny$\substack{+3.3 \\ -1.9}$}} & 0.22 & 6.3{\raisebox{0.5ex}{\tiny$\substack{+4.5 \\ -2.5}$}} \\

    0.31 & 9.3{\raisebox{0.5ex}{\tiny$\substack{+4.8 \\ -4.2}$}} & 0.28 & 1.0{\raisebox{0.5ex}{\tiny$\substack{+3.2 \\ -0.8}$}} \\
    
    \cline{1-2} \multicolumn{2}{c|}{EPTA~\cite{EPTA:2023xxk}} & 0.33 & 6.3{\raisebox{0.5ex}{\tiny$\substack{+4.4 \\ -4.0}$}} \\

     0.097 & 8.0{\raisebox{0.5ex}{\tiny$\substack{+4.0 \\ -2.7}$}} & 0.39 & 3.7{\raisebox{0.5ex}{\tiny$\substack{+3.7 \\ -1.5}$}} \\

     0.19 & 9.6{\raisebox{0.5ex}{\tiny$\substack{+2.9 \\ -1.9}$}} & 0.44 & 7.1{\raisebox{0.5ex}{\tiny$\substack{+3.9 \\ -2.6}$}} \\

     0.29 & 8.2{\raisebox{0.5ex}{\tiny$\substack{+3.8 \\ -8.2}$}} & 0.50 & 1.6{\raisebox{0.5ex}{\tiny$\substack{+3.2 \\ -1.0}$}} \\

     0.39 & 11.0{\raisebox{0.5ex}{\tiny$\substack{+4.5 \\ -2.9}$}} & 0.55 & 3.6{\raisebox{0.5ex}{\tiny$\substack{+3.6 \\ -2.09}$}} \\

     0.48 & 5.2{\raisebox{0.5ex}{\tiny$\substack{+10.0 \\ -5.5}$}} & & \\

     0.58 & 0.6{\raisebox{0.5ex}{\tiny$\substack{+4.8 \\ -0.4}$}} & &
\end{tabular}
\caption{Data points from the PTA characteristic spectrum used in our fits. We neglect correlations between different frequency bins and/or experiments.}
\label{tab:data_points}
\end{table}

\end{appendix}
\bibliography{ref}
\bibliographystyle{utphys}

\end{document}